\documentclass{article}

\usepackage{PRIMEarxiv}

\usepackage[utf8]{inputenc} 
\usepackage[T1]{fontenc}    
\usepackage{hyperref}       
\usepackage{url}            
\usepackage{booktabs}       
\usepackage{amsfonts}       
\usepackage{nicefrac}       
\usepackage{microtype}      
\usepackage{lipsum}
\usepackage{fancyhdr}       
\usepackage{graphicx}       
\graphicspath{{media/}}     

\usepackage{amsmath,epsfig}

\usepackage{amssymb}

\usepackage{multirow}
\usepackage{multicol}
\usepackage{threeparttable}
\usepackage{mathrsfs}
\usepackage{graphicx}
\usepackage{subcaption}
\usepackage{booktabs}

\title{Cognitive Effort Measures Driven by Fixation Induced Retinal Flow in Visual Scanning Behavior during Virtual Driving
}

\author{
  Runlin Zhang, Qing Xu \\
  College of Intelligence and Computing \\
  Tianjin University \\
  Tianjin\\
  \texttt{\{runlin, qingxu\}@tju.edu.cn} \\
   \And
  Simon Parkinson \\
  School of Computing and Engineering \\
  University of Huddersfield \\
  Huddersfield\\
  \texttt{S.Parkinson@hud.ac.uk} \\
   \AND
  Klaus Schoeffmann \\
  Institute of Information Technology \\
  Alpen-Adria Universitat Klagenfurt \\
  Klagenfurt\\
  \texttt{ks@itec.aau.at} \\
   \And
  Yu Chen \\
  School of Foreign Languages \\
  Southeast University \\
  Nanjing\\
}

\begin{document}
\maketitle

\begin{abstract}

 In this paper, we consider the problem of visual scanning mechanism underpinning sensorimotor tasks, such as walking and driving, in dynamic environments. We exploit eye tracking data for offering two new cognitive effort measures in visual scanning behavior of virtual driving. By utilizing the retinal flow induced by fixation, two novel measures of cognitive effort are proposed through the importance of grids in the viewing plane and the concept of information quantity, respectively. With psychophysical studies, two proposed cognitive effort measures have shown their significant correlation with widely used objective measurements of cognitive effort. Our results suggest that the quantitative exploitation of eye tracking data provides an effective approach for the evaluation of sensorimotor activities.
\end{abstract}


\section{Introduction}
\label{sec:Intro}

Visual scanning and eye tracking are important for the living of any human in natural surroundings~\cite{ghosh2023automatic,eyetracking}. Visual scanning is indeed the foundation for a human to perform common and everyday sensorimotor tasks, such as walking and driving. Actually, the understanding of the mechanism behind visual scanning has been valuable since late 1970 and, is especially helpful and beneficial to making stark and essential progress in both theoretical and practical perspectives~\cite{shiferaw2019review}.

Basically, it is significant to make clear how much a human can achieve for sampling visual information through visual scanning, bearing one of the most fundamental points involved in the understanding of visual scanning mechanism \cite{shiferaw2019review}. Cognitive effort is an important approach to comprehending cognitive control and visual scanning behavior~\cite{WPS2018}. Actually, cognitive effort basically depicts subjective engagement for assessing the human’s internal state during tasks~\cite{WB2015}. Thus, cognitive effort plays an important role in visual scanning and visuo-motor behavior, but how to measure and in particular how to objectively assess the cognitive effort have been a paramount concern in both theory and practice~\cite{WPS2018,Luca2022}.

It is noted that the visual motion, which always occurs during the visual scanning behavior of sensorimotor tasks~\cite{negahdaripour1998revised}, has been rarely touched in the cognitive effort measure. In this paper, based on the so-called fixation induced retinal flow~\cite{LvXu2021} that is a quantitative description for the visual motion in the visual interactive perception on environmental stimuli, the importance of grids in the viewing plane is developed. A cognitive effort measure, called the view importance based cognitive effort measure ($CEM_{VI}$), is proposed, through employing \emph{Shannon} entropy based complexity \cite{cover2012elements} of the probability distribution of the view importance of grids. Still, based on the fixation induced retinal flow, the amount of perception of the visual motions during sensorimotor tasks is obtained using the square root of \emph{Jensen-Shannon divergence}, which is a true mathematical metric~\cite{lin1991divergence}. Then, in terms of the concept of information quantity~\cite{cover2012elements}, the perceived amount of visual motions is transformed to be utilized for satisfactorily defining the information quantity based cognitive effort measure ($CEM_{IQ}$) to understand the cognitive status of humans during driving tasks. To the best of our knowledge, this is the first time, based on the exploitation of visual motion and a true mathematical metric, to effectively define the quantitative and objective evaluations of the classical and subjective cognitive effort. Our proposal paves a novel path for behaviometric discovery by the utilization of eye tracking data.

\section{Related Works}
\label{sec:related}

An initial consideration for discussing the measurement of the visual scanning behavior is to select suitable eye tracking indices \cite{eyetracking} naturally associated with cognitive processing. Fixation and saccade \cite{eyetracking} are classic indices for this purpose. But, direct and indirect usages of these indices (for examples, rate/duration of them and their simple combinations) are more applicable in specific application scenarios, rather than in the general evaluation of visual scanning behavior~\cite{jeong2019driver}. In addition, pupil dilation and blink rate are two widely used eye tracking indices for the study of cognition and psychology \cite{eckstein2017beyond}. Notice specifically that eye tracking has been expected as a strong estimator for task performance in many professional fields \cite{2020Amphetamine}.

Basically, the knowing of cognitive effort plays a significant character in the procedure of all kinds of cognitive processing~\cite{WB2015,Luca2022,Shenhav2017}. The cognitive effort, which has started within educational psychology, is a classic measure for subjective engagement~\cite{WB2015}. From the perspective of cost-benefit decision-making, the cognitive effort is deemed as an amplitude or intensity of behavior in the fulfillment of cognitive control for accomplishing tasks~\cite{Shenhav2017}. The assessment of cognitive effort, which is used for the estimation of the human’s internal state, has been largely encouraged in the area of ergonomics and human factors \cite{Luca2022}. And undoubtedly, the measurement of cognitive effort in the domain of visual scanning and visuo-motor has attracted a lot of attention in both theory and practice~\cite{WPS2018}. But, the visual motion, which usually appears during sensorimotor activities, has not been used in the measurement of cognitive effort. In addition, evaluating the cognitive effort, particularly in an objective way, bears a big challenge \cite{Luca2022}.

The visual motion perceived during a fixation in a sensorimotor task, which is the so-called fixation induced retinal flow in this paper, has been introduced based on the concepts of eye tracking and optical flow in the literature of visual scanning and visuo-motor~\cite{LvXu2021}. Considering that the fixation induced retinal flow is very important as the fundamental basis for establishing the two proposed measures, its methodology is specially depicted in Section~\ref{sec:retinalflow}.

Additionally, previous research has demonstrated numerous eye movement measures related to cognitive effort in humans during tasks. For example, when a driver's cognitive load increases, the driver's periphery/mirror/instrument check rate (hereafter referred to as check rate) tends to decrease~\cite{he2022classification}. Stationary Gaze Entropy ($SGE$) is used to measure the level of fixation dispersion during the eye scanning process~\cite{shiferaw2019review}. Shiferaw et al. found that during driving, if the driver is in an abnormal state such as hungover or fatigued, their $SGE$ shows significant changes~\cite{shiferaw2019review}. Entropy rate, a concept in information theory, is used to describe the rate at which a random process generates information. In more specific applications, it can quantify the uncertainty and complexity of a signal or data sequence. It has also been confirmed to change with variations in cognitive load~\cite{tong2023measuring}.

\section{Methods}
\subsection{The Retinal Flow induced by Fixation and Visual Scanning Efficiency}
\label{sec:retinalflow}

The retinal flow induced by a fixation is introduced based on the identification of the visual motion resulted from a fixated stimulus~\cite{LvXu2021}.

\begin{figure}[h]
  \centering
  \centerline{\includegraphics[width=0.5\linewidth]{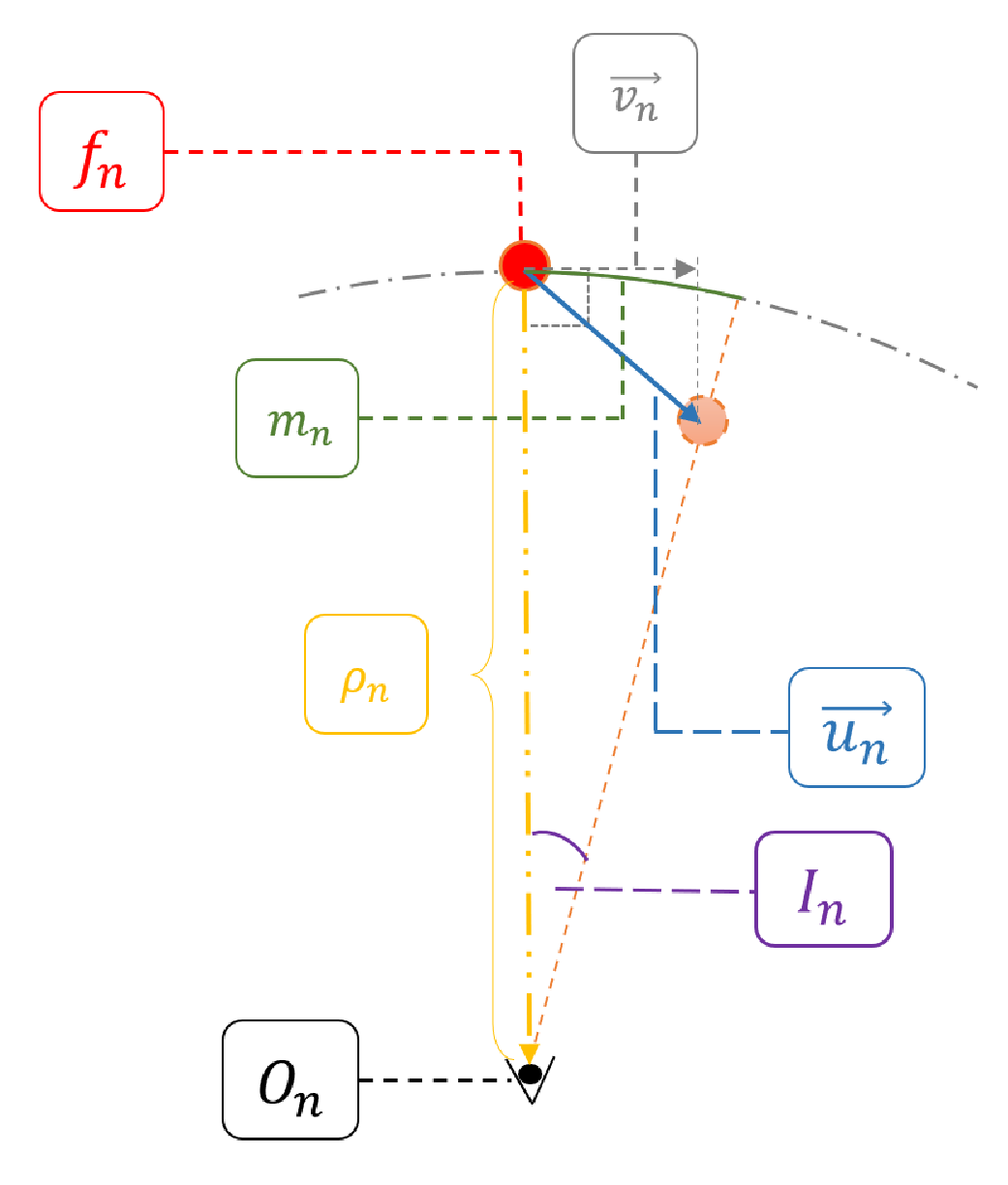}}
\caption{An illustration of retinal flow $I_n$ induced by fixation $f_n$.}
\label{fig:TotelVelocity}
\end{figure}

A fixation $f_n$ (its index is $n$) with duration $\tau_n$ by an observer $O$ in a \emph{3D} environment is shown in Fig.~\ref{fig:TotelVelocity}. Here, $\rho_n$ is the distance between $O$ and $f_n$ in the direction of viewing, and $\rho_n$ is the depth of $f_n$ from the perspective of $O$. Considering, during $\tau_n$, there is a relative motion displacement $\vec{u_n}$ happened to $f_n$, and $\vec{\upsilon_n}$ is taken as the optical flow vector \cite{negahdaripour1998revised} for the fixated stimulus by $f_n$ . Actually, $\vec{\upsilon_n}$ is the projection of $\vec{u_n}$ in the direction of optical flow vector, and thus perpendicular to the direction of viewing. For computation simplicity, the length of trajectory segment of the circular motion of $f_n$ centered on $O$, $m_n$, acts as an approximated magnitude of $\vec{\upsilon_n}$. The central angle $I_n$ subtended by $m_n$,
\vspace{-0.1cm}
\begin{equation}
\label{eq:I}
  I_n = \dfrac{m_n}{\rho_n},
\end{equation}
\vspace{-0.4cm}

\noindent is further used to define a perceived magnitude of optical flow, for characterizing the amount of visual motion perceived by the observer during the fixation $f_n$. $I_n$ is called as the fixation induced retinal flow in this paper, because this quantity represents the amount of perceived optical flow in the course of a fixation. The definition of $I_n$ meets the usual practice that, angle is widely used as the representation of magnitude or amount in eye tracking~\cite{eyetracking}. Note $I_n$ explicitly uses the depth cue $\rho_n$, enabling the fixation induced retinal flow to convey this important and special cue in \emph{3D} environments.

Visual scanning is done by a performer through a sequence of fixations, so that meeting the requirement of visual sampling the surrounding environments for fulfillment of a sensorimotor task. The probability distribution of fixations $P_f$, namely the fixation distribution, which is built up based on the normalized histogram of fixation locations in a \emph{3D} environment, is used for a representation of the visual scanning behavior. The fixation induced retinal flow is used to construct the retinal flow probability distribution $P_r$. Previous work measured the difference between $P_f$ and $P_r$ by the Square Root of Jensen-Shannon Divergence (SRJSD) between them, 
\begin{equation}
\label{eq:vse}
    SRJSD(P_f||P_r)  =  \sqrt{JSD(P_f||P_r)},
\end{equation}
for assessing the so-called visual scanning efficiency~\cite{LvXu2021}. $SRJSD(P_f||P_r)$ plays a basic function for understanding the mechanism of visual scanning behavior.

\subsection{The Proposed Cognitive Effort Measure Based on the View Importance}
\label{sec:impivse}

During a sensorimotor task such as driving, for the purpose of safety and stability, the driver usually focuses varied attention on regions in the viewing plane. For example, central and peripheral viewing regions are paid large and small attention and/or importance, respectively, to achieve stable driving, if the central viewing regions dominate the road for driving. That is to say, the importance of stimulus observation plays an important factor in performing the driving tasks, considering that a stimulus in 3D environment corresponds to at least a region in the viewing plane.

In this paper, the amount of the perceived visual motion resulted from a fixated stimulus, which is characterized as the fixation induced retinal flow, is utilized to define
\vspace{-0.1cm}
\begin{equation}
\label{eq:stiimp}
  J_n = \dfrac{I_n}{|\vec{u_n}|}
\end{equation}
\vspace{-0.4cm}

\noindent as the importance of stimulus observation. Notice that the observation importance $J_n$ takes into consideration of the motion displacement $|\vec{u_n}|$ of a fixation during its duration, for explicitly signaling the influence of the eye yaw rotation on the observation of the fixated stimulus. This means that when $J_n$ is larger the obtainment of fixation is easier, and conversely, when $J_n$ is smaller the obtainment of fixation is more challenging. That is, from the perspective of stimulus observation, $J_n$ provides a kind of indicator for pointing out how much effort should be exerted to observe and perceive a stimulus. From the viewpoint of a stimulus itself, $J_n$ offers a measurement of its importance for observation and perception.
And then, the view importance of a region in the viewing plane is obtained, by accumulating all the values of the observation importance of the corresponding stimulus to this region. A region corresponds to a single element of $N_g \times N_g$ grids in the viewing plane (currently, $N_g = 5$ achieves good results in this paper, and other options on $N_g$ will be done as a future work). The normalized histogram of values of the view importance of grids is created, to obtain a probability distribution $P_{g}=\{p_{g}(j)|j=1,\cdots,N_g\}$ of the view importance of grids. The \emph{Shannon} entropy 
\vspace{-0.2cm}
\begin{equation}
\label{eq:cemimp}
  CEM_{VI}= -\sum\limits_{j=1}^{N_g}p_g(j)\log_{2}p_g(j)
\end{equation}
\vspace{-0.5cm}

\noindent of this probability distribution is proposed to evaluate the degree of balance for visually scanning various grids in the viewing plane, leading to the view importance based cognitive effort measure (${CEM_{VI}}$) in our 
paper. This entropy based complexity evaluation on the view importance of grids, which indeed takes into account of the non-trivial interaction between the observations on various grids. As a result, this proposed ${CEM_{VI}}$ indicates the degree of a systematic perception of all the visual motions during the sensorimotor driving, as well suggests an intensity or amplitude of the visual scanning behavior and behaves as an assessment function for cognitive effort.

\subsection{The Information Quantity of Perceived Visual Motion and the Corresponding Proposed Cognitive Effort Measure}
\label{sec:CEM}

The developed measure of the amount of perceived visual motions, $SRJSD(P_{f}||P_{r})$, can be studied from the perspective of probability and information theory~\cite{cover2012elements}. That is, $SRJSD(P_{f}||P_{r})$, in fact, can be considered as a probability $p$ an event occurs, because it ranges from 0 to 1~\cite{lin1991divergence}. As a result, $SRJSD(P_{f}||P_{r})$ gives a probability of perception of the visual motions. According to information theory, the logarithmic probability of occurrence ($-\log_2 p$) represents the quantity of information conveyed by the occurrence~\cite{cover2012elements}. It is obvious that the information quantity $-\log_2 SRJSD(P_{f}||P_{r})$ indicates the quantified amount of perception of all the visual motions during a sensorimotor task. Notably, the amplitude for the perception of visual motions and for the completion of sensorimotor tasks, reflects the meaning of cognitive effort~\cite{Shenhav2017}. Thus the logarithm transformation of perceived visual motion is taken as the core function for the proposed information quantity based cognitive effort measure ($CEM_{IQ}$).




Because the distributions of fixations and fixation transitions reflect different aspects of visual scanning, a combination of both these two distributions should provide a more complete understanding of visual scanning behavior, as has been pointed out in relevant work~\cite{shiferaw2019review}. Indeed, the performer of visual scanning voluntarily exerts some cognitive effort to do a
unidirectional switch between two neighboring fixations, and the visual motion induced by the first fixation perceived/cognized in the procedure of completing one fixation transition measures this effort. Therefore, we propose to utilize the fixation transition distribution $P_{fs}$ and the retinal flow distribution $P_{rs}$ based on fixation transition to obtain the definition of $CEM_{IQ}$. The approach to obtaining $P_{fs}$ is similar to that of the fixation distribution $P_{f}$, but here the fixation sequence is employed. Correspondingly, $P_{rs}$ can be easily obtained based on $P_{fs}$. $-\log_2 SRJSD(P_{fs}||P_{rs})$ gives an information quantity for the perceived visual motion during a single fixation transition. We define $CEM_{IQ}$ by the division between these two information quantities, $-\log_2 SRJSD(P_{f}||P_{r})$ and $-\log_2 SRJSD(P_{fs}||P_{rs})$, as
\begin{equation}
\label{eq:NE}
\begin{aligned}
    CEM_{IQ} = \dfrac{-\log_2 SRJSD(P_f||P_r)}{-\log_2 SRJSD(P_{fs}||P_{rs})},
\end{aligned}
\end{equation}
\noindent for characterizing a cognitive effort during a sensorimotor task.


\begin{figure}[h]
  \centering
  \centerline{\includegraphics[width=0.5\linewidth]{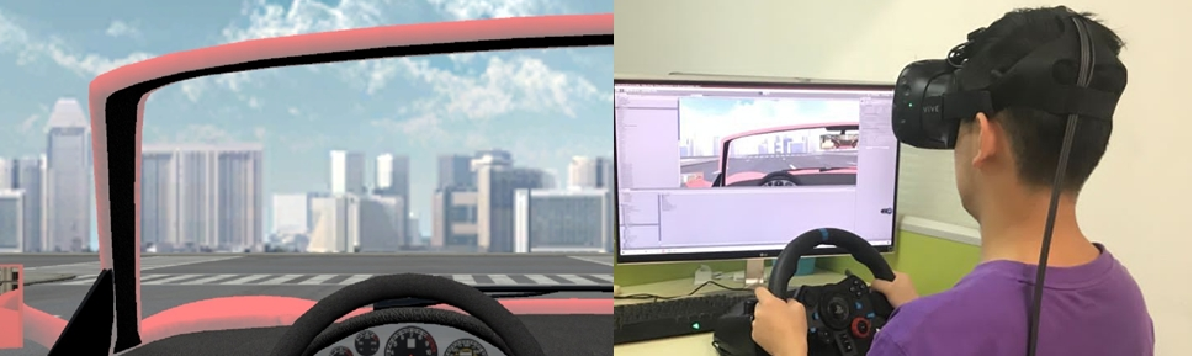}}
\vspace{-0.2cm}
\caption{A participant is performing a driving task in virtual reality.}
\label{fig:VR}
\end{figure}

\vspace{-0.3cm}
\section{Experiment}
\label{sec:experiment}
\vspace{-0.1cm}
\subsection{Participants}
\vspace{-0.1cm}
14 Master/Phd students ($5$ females; age range: $21$-$29$, Mean = $21.3$, SD = $2.37$) with driving experience (they hold their driver license at least one and a half years) from our University volunteer to participate in the psychophysical studies. All of the participants have normal/corrected-to-normal visual acuity and normal color vision. There is no participant having adverse reaction to the virtual environment we set up for the studies.

\vspace{-0.1cm}
\label{subsec:VE}
\vspace{-0.1cm}


\vspace{-0.1cm}
\subsection{Apparatus}
\vspace{-0.1cm}
\emph{HTC Vive} headset is used to display the vitural environment for participants. The eye-tracking equipment is \emph{7INVENSUN Instrument aGlass DK\uppercase\expandafter{\romannumeral2}}, which is embedded into the \emph{HTC Vive} display to capture visual scanning data in a frequency of $90$ Hz and in an accuracy of gaze position of $0.5^\circ$. The driving device is a \emph{Logitech G29} steering wheel. Participants listen the ambient traffic and car engine sounds in \emph{VE} by speakers. The visual and driving behaviors of participants are displayed on desktop monitor for observation.
\vspace{-0.1cm}
\subsection{Driving Task}
\vspace{-0.1cm}
As discussed in Section \ref{subsec:VE}, the task directed focus on visual scanning and driving is taken. And as a result, participants are required to keep driving at a target speed of $40$ km/h, for speed control. This paper takes the inverse of the mean acceleration of vehicle to denote the driving performance. The smaller the mean acceleration, the higher the driving performance becomes, and \emph{vice versa}. In fact, this kind of performance measure has been used a lot in literature \cite{yadav2019effect}. An example of performing driving tasks is presented in Fig.~\ref{fig:VR}.

\subsection{Procedure}

Each participant completes 4 test sessions with the same task requirements and the same driving routes, with a 9-point calibration for the eye tracker at the beginning of each session and with an interval of one week between every two sessions. Data for visual scanning and driving behaviors are recorded during test sessions. In this paper, a trial represents a test session, and there are $14*4=56$ valid trials in all (obviously this gets a large enough sample size~\cite{lehmann1999elements}). A preparation session is applied to participants before each test session to let them know the purpose and procedure about the studies.

\renewcommand{\arraystretch}{1.2} 
\begin{table*}[t]
  \centering
  \fontsize{6.3}{8.2}\selectfont
  \begin{threeparttable}
  \caption{Correlation between the proposed measures and pupil size change/fixaion rate}
  \label{tab:correlNBE}
    \begin{tabular}{ccccccccccccc}
    \toprule
    \multirow{3}{*}{Correlation}&\multicolumn{6}{c}{Pupil Size Change}&\multicolumn{6}{c}{Fixaion Rate}\cr
    \cline{2-13}
     &\multicolumn{2}{c}{Pearson}&\multicolumn{2}{c}{Kendall}&\multicolumn{2}{c}{Spearman}&\multicolumn{2}{c}{Pearson}&\multicolumn{2}{c}{Kendall}&\multicolumn{2}{c}{Spearman} \cr
    \cline{2-13}
    &CC&p-value&CC&p-value&CC&p-value&CC&p-value&CC&p-value&CC&p-value\cr
    \midrule
      $CEM_{VI}$&0.38&$p^{**}<$0.01&0.27&$p^{**}<$ 0.01&0.39&$p^{**}<$ 0.01  &-0.19&$p>$0.05&-0.04&$p>$0.05&-0.04&$p>$0.05\cr
      $CEM_{IQ}$&0.27&$p^{*}<$0.05&0.20&$p^{*}<$ 0.05&0.27&$p^{*}<$ 0.05 &-0.46&$p^{***}<$0.001&-0.23&$p^{*}<$ 0.05&-0.36&$p^{**}<$ 0.01\cr
      $Check Rate$&0.15&$p>$0.05&0.14&$p>$0.05&0.21&$p>$0.05 &0.35&$p^{**}<$0.01&0.32&$p^{**}<$ 0.01&0.46&$p^{**}<$ 0.01\cr
      $SGE$&0.03&$p>$0.05&0.02&$p>$0.05&0.05&$p>$0.05 &0.01&$p>$0.05&-0.02&$p>$0.05&-0.04&$p>$0.05\cr
      $Entropy Rate$&-0.02&$p>$0.05&0.01&$p>$0.05&0.03&$p>$0.05 &-0.07&$p>$0.05&-0.10&$p>$0.05&-0.18&$p>$0.05\cr
    \bottomrule
    \end{tabular}
    \end{threeparttable}
\end{table*}

\section{Results and Discussions}
\label{sec:resultsDis}

\vspace{-0.15cm}
\subsection{Correlation Results}
\vspace{-0.05cm}
\label{subsec:NEresult}
As widely used in literature as a measure of cognitive effort~\cite{eckstein2017beyond,joshi2016relationships}, pupil size change has been accepted as an autonomic and reflexive measure of cognitive effort. In this paper, the standard deviation of pupil size \cite{chen2011eye,WPS2018} during each trial is utilized to represent the pupil size change because of its simplicity and effectiveness. The fixation rate is also used to measure cognitive effort~\cite{souchet2022measuring}, because corresponding studies have indicated that factors influencing pupil size are not solely due to cognitive effort~\cite{petersch2022gaze}. It is clear that cognitive load affects both human pupillary response and fixation based eye movements. Therefore, both pupil size change and fixation rate are taken in this paper as the definitive quantitative ground truth for cognitive effort. Three classic quantitative measures of cognitive effort, check rate~\cite{he2022classification}, $SGE$~\cite{shiferaw2019review} and entropy rate~\cite{tong2023measuring} are used as comparison for evaluating the effectiveness of our proposed measures.

\begin{figure}[h]
  \centering
  \centerline{\includegraphics[width=1\linewidth]{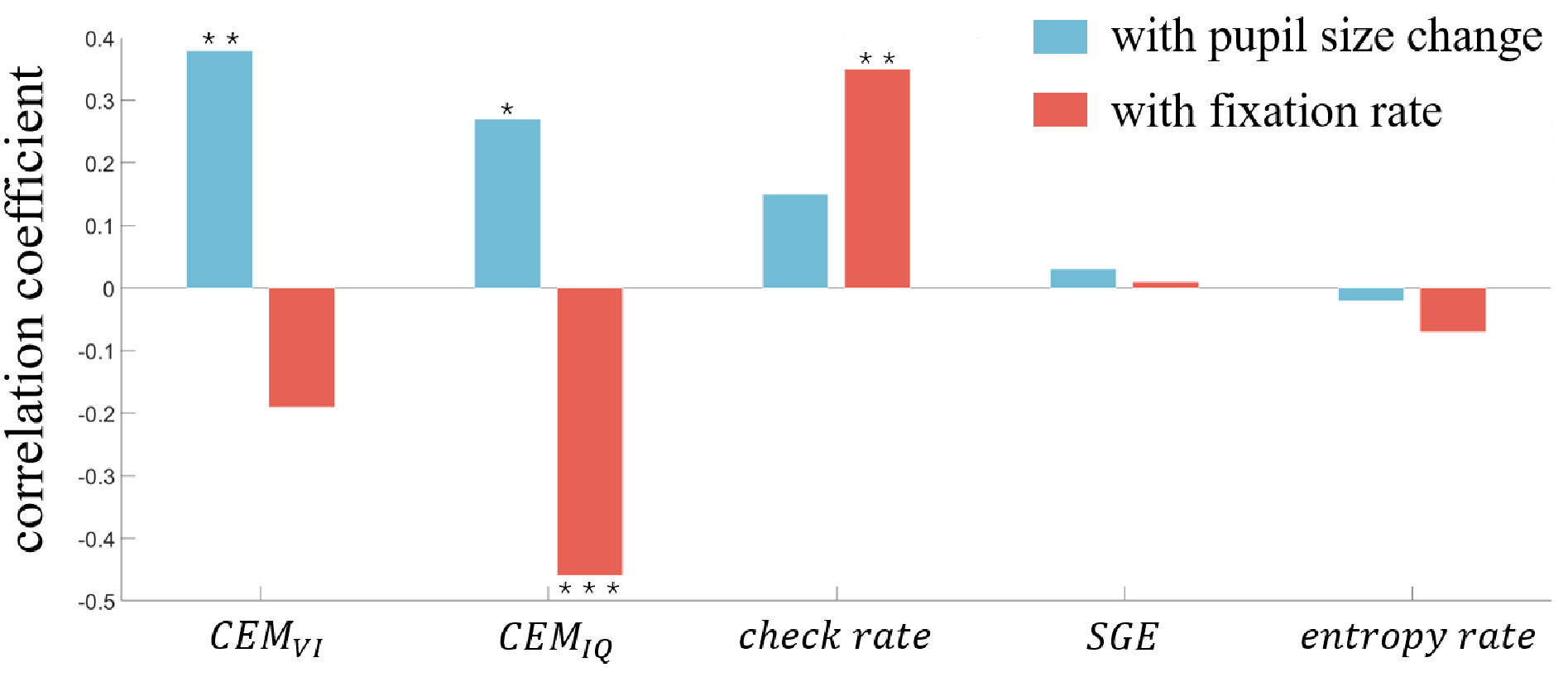}}
\caption{Correlation coefficients of eye movement measures with pupil size change/fixation rate (*: $p<0.05$, **: $p<0.01$, ***: $p<0.001$)}
\label{fig:cc_res}
\end{figure}

We validate the relationship between two proposed measures and pupil size change/fixation rate through three commonly used correlation coefficients called Pearson Linear Correlation Coefficient (PLCC), Kendall Rank Order Correlation Coefficient (KROCC) and Spearman Rank Order Correlation Coefficient (SROCC). The correlation results are listed in Table~\ref{tab:correlNBE}. We find that, $CEM_{IQ}$ shows a significant correlation with both pupil size change and fixation rate. $CEM_{VI}$ exhibits a significant correlation with pupil size change. The check rate is not related to pupil size change, yet it has a significant correlation with fixation rate. But, $SGE$ and entropy rate are not correlated with pupil size change and fixation rate. Correlation analysis between eye movement measures
with pupil size change/fixation rate is also clearly shown in Fig.~\ref{fig:cc_res}.

\subsection{General Discussions}
\vspace{-0.05cm}
\label{subsec:discussion}

The proposed cognitive effort measures are based on the methodology of information theory, through taking advantage of the perceived visual motion always happening in dynamic environments during sensorimotor tasks. In fact, our proposal actually satisfies the definition of cognitive effort in terms of information processing~\cite{Shenhav2017}. The significant positive correlation between $CEM_{VI}$ and pupil size change suggests the principle that the more chaotic and varied the distribution of the importance of different areas in the driver's viewing plane, the higher the corresponding cognitive effort on the driver. Indeed $CEM_{VI}$ is designed based on this principle to assess the driver's cognitive load. Overall, $CEM_{IQ}$ achieves the best in measuring cognitive effort, from both pupillary and fixation perspectives. From the viewpoints of both pupil size and fixation, a consistent conclusion can be drawn: the larger the $CEM_{IQ}$, the greater the cognitive effort. The significant correlation between $CEM_{IQ}$ and the ground truth of cognitive effort demonstrates that the higher the proportion of perceived visual motion information among all perceived potential eye movement changes, the higher the cognitive effort on the driver. The achievement by $CEM_{IQ}$ is also evidenced by a comparison between it with other classic measurements of cognitive effort based on eye movement. Among the three measures under comparison, only the check rate has a significant correlation with the fixation rate. We believe this is because the check rate itself is specifically related to the fixation distribution for driving. In a word, we consider our proposed $CEM_{IQ}$ to have the best robustness, being applicable in a broader range of scenarios and potentially yielding a more accurate measurement of cognitive effort. In the meantime, we believe that the definition of cognitive effort in terms of the information quantity and of ``physics" is worthwhile, and we will continue deep exploitation in this avenue.

Considering that we have made progress on the exploitation of eye tracking data, as a behaviometric, for the evaluations on cognitive effort, a further investigation into the relationship between these two proposed measures and the performance of sensorimotor tasks will be done in future work. And actually, this could be a working path illuminated based on the exploitation of Yerkes-Dodson law~\cite{watters1997caffeine}.

Notice that the findings of this paper may not be applicable for all cases, but it does work in the context of our topic. Due to that the visual scanning and visuo-motor behavior is exceptional important in virtual and real-world sensorimotor tasks, what we have achieved on the measurement of cognitive effort in virtual driving should be potentially helpful for ergonomic evaluation pragmatically, in many practical and relevant applications.

\section{Conclusions and Future Works}
\label{sec:Con}
In this paper, we take an important step for thorough understanding the mechanism of visual scanning in virtual driving. This paper has established, in an objective and quantitative way, two new measures for the subjective cognitive effort, mainly by utilizing information theoretic tools. Our proposal is well done through a methodology that exploits the perceived visual motions in a sensorimotor task. As far as we can know, no research up to now has reported this kind of finding to shed light on the issue of cognitive effort measure for the visual scanning behavior during sensorimotor tasks. Additionally, the proposed cognitive effort measures may offer a new perspective on the inherent relationship between task directed visual scanning and eye tracking data, so as to help the development of behaviometric discovery, from both theoretical and practical perspectives.

In the near future, we will investigate our proposed methodology and measures for real-life driving scenarios, for instance, for crash risk problem \cite{Victor2015}. In consideration of the critical role of illumination conditions for driving, we will exploit the manipulation of illumination levels in a detailed quantitative way, to comprehensively understand the mechanism of cognitive effort during visual scanning behavior. Also, physiological signals such as heartbeat
~\cite{Active2020Galvez} will be investigated for understanding the relationships and interplays between these signals and eye tracking data, for the sake of cognitive effort.

\bibliographystyle{unsrt}  
\bibliography{references}

\end{document}